\documentstyle[prd,preprint,tighten,aps,eqsecnum,amssymb,newlfont,epsfig]{revtex}
\begin{document}
\draft
\preprint{
\begin{tabular}{r}
\null \vspace{1cm} \null\\
KIAS-P98046\\
DFTT 70/98\\
hep-ph/9902261\\
\null \vspace{1cm} \null
\end{tabular}
}
\title{Implications of Super-Kamiokande atmospheric low-energy data
for solar neutrino oscillations}
\author{
C. Giunti$^{\mathrm{a,b}}$,
C.W. Kim$^{\mathrm{a,c}}$,
U.W. Lee$^{\mathrm{a,d}}$
and
V.A. Naumov$^{\mathrm{a,e}}$}
\address{
\begin{tabular}{c}
$^{\mathrm{a}}$
School of Physics,
Korea Institute for Advanced Study,
Seoul 130-012, Korea
\\
$^{\mathrm{b}}$
INFN, Sez. di Torino, and Dip. di Fisica Teorica,
Univ. di Torino,
I--10125 Torino, Italy
\\
$^{\mathrm{c}}$
Dept. of Physics $\&$ Astronomy,
The Johns Hopkins University,
Baltimore, MD 21218, USA
\\
$^{\mathrm{d}}$
Department of Physics,
Mokpo National University,
Chonnam 534-729,
Republic of Korea
\\
$^{\mathrm{e}}$
Irkutsk State University, Gagarin Boulevard 20,
RU-664003 Irkutsk, Russia
\end{tabular}
}
\maketitle
\begin{abstract}
It is shown that the high-$\Delta{m}^2$ part of
the large mixing angle MSW solution of the solar neutrino problem
is disfavored by the Super-Kamiokande data
for low-energy upward-going events.
A quantitative bound is obtained in the three-neutrino scheme
with a negligibly small element $U_{e3}$
of the neutrino mixing matrix,
as indicated by the result of the CHOOZ long-baseline
$\bar\nu_e\to\bar\nu_e$ oscillation experiment.
\end{abstract}


\section{Introduction}
\label{Introduction}

The recent results of the Super-Kamiokande atmospheric neutrino experiment
\cite{SK-atm-98}
provided a model-independent evidence in favor of neutrino oscillations
that has been searched for since the discovery of the theory of neutrino oscillations
\cite{discovery-of-neutrino-oscillations}.
The solar neutrino problem
\cite{solar-neutrino-problem}
and the atmospheric neutrino anomaly
can be explained by neutrino oscillations in the scheme with mixing
of the three flavor neutrinos $\nu_e$, $\nu_\mu$ and $\nu_\tau$
with three massive neutrinos $\nu_1$, $\nu_2$ and $\nu_3$.
In this case the oscillations of solar and atmospheric
neutrinos are due to the mass-squared differences
$ \Delta{m}^2_{21} \equiv m_2^2 - m_1^2 $
and
$ \Delta{m}^2_{31} \equiv m_3^2 - m_1^2 $,
respectively,
where $m_k$ is the mass of the massive neutrino $\nu_k$
($k=1,2,3$),
and
\begin{equation}
\Delta{m}^2_{21}
\ll
\Delta{m}^2_{31}
\,.
\label{001}
\end{equation}

The atmospheric neutrino anomaly has been observed in the
Kamiokande \cite{Kam-atm-94},
IMB \cite{IMB95},
Soudan \cite{Soudan97},
Super-Kamiokande \cite{SK-atm-98}
and MACRO \cite{MACRO-98} experiments.
The fit of the high-statistics
Super-Kamiokande atmospheric neutrino data with
two-neutrino
$\nu_\mu\to\nu_\tau$
oscillations yielded an allowed region
in the $\sin^22\vartheta_{\mathrm{atm}}$--$\Delta{m}^2_{31}$
plane delimited by \cite{SK-atm-98}
\begin{equation}
\sin^22\vartheta_{\mathrm{atm}}
\gtrsim
0.7
\,,
\qquad
3 \times 10^{-4}
\, \mathrm{eV}^2
\lesssim
\Delta{m}^2_{31}
\lesssim
9 \times 10^{-3}
\, \mathrm{eV}^2
\,,
\label{401}
\end{equation}
at 99\% CL.
Furthermore,
the results of the
Kamiokande \cite{Kam-atm-94},
IMB \cite{IMB95}
and
Soudan \cite{Soudan97}
experiments,
together with the preliminary new data of the Super-Kamiokande experiment
\cite{Suzuki-WIN99},
indicate that
$\Delta{m}^2_{31}$
is larger than a few times
$ 10^{-3} \, \mathrm{eV}^2 $.
Hence,
in the following we will consider
\begin{equation}
2 \times 10^{-3}
\, \mathrm{eV}^2
\lesssim
\Delta{m}^2_{31}
\lesssim
9 \times 10^{-3}
\, \mathrm{eV}^2
\,.
\label{406}
\end{equation}

The rates of the solar neutrino experiments
(Homestake \cite{Homestake98},
Kamiokande \cite{Kam-sun-96},
GALLEX \cite{GALLEX96},
SAGE \cite{SAGE96}
and
Super-Kamiokande \cite{SK-sun-98-PRL,SK-sun-nu98})
can be explained
by $\nu_e\to\nu_\mu$ and/or  $\nu_e\to\nu_\tau$
oscillations in vacuum
or MSW \cite{MSW} resonant transitions in the interior of the sun
with a small or a large mixing angle $\vartheta_{\mathrm{sun}}$.
Here we are interested in the large mixing angle MSW solution
of the solar neutrino problem whose allowed region in the
two-neutrino mixing
parameter space
$\sin^22\vartheta_{\mathrm{sun}}$--$\Delta{m}^2_{21}$
is delimited by
\cite{Bahcall-Krastev-Smirnov98}
\begin{equation}
\sin^22\vartheta_{\mathrm{sun}}
\gtrsim
0.5
\,,
\qquad
6 \times 10^{-6}
\, \mathrm{eV}^2
\lesssim
\Delta{m}^2_{21}
\lesssim
3 \times 10^{-4}
\, \mathrm{eV}^2
\,,
\label{402}
\end{equation}
at 99\% CL.
These large values of the mass-squared difference
$\Delta{m}^2_{21}$
and
mixing angle
$\sin^22\vartheta_{\mathrm{sun}}$
are large enough to
have an effect on upward-going low-energy atmospheric neutrinos.
Indeed,
these neutrinos
have energy
$ 100 \, \mathrm{MeV} \lesssim E \lesssim 1 \, \mathrm{GeV} $
and propagate for a distance
$ 10^3\, \mathrm{km} \lesssim L \lesssim 10^4 \, \mathrm{km} $.
This means that
the corresponding energy-dependent phase for neutrino oscillations is
\begin{equation}
\frac{ \Delta{m}^2_{21} L }{ 2 E } \sim 1
\label{403}
\end{equation}
and hence must be taken into account in the analysis of atmospheric neutrino data
\cite{KimLee98}.

If
the disappearance of atmospheric $\nu_\mu$'s observed by the Super-Kamiokande
and other atmospheric neutrino experiments
is due to $\nu_\mu\to\nu_\tau$ oscillations driven by the mass-squared difference
$\Delta{m}^2_{31}$,
the high-$\Delta{m}^2_{21}$ part of
the large mixing angle MSW solution of the solar neutrino problem
is disfavored by the Super-Kamiokande data relative to low-energy
upward-going $e$-like and $\mu$-like events.
Indeed,
Eq.(\ref{403}) implies that
low-energy
upward-going
neutrinos undergo not only $\nu_\mu\to\nu_\tau$
transitions driven by $\Delta{m}^2_{31}$
but also
$\nu_\mu\leftrightarrows\nu_e$,
$\nu_\mu\to\nu_\tau$
and
$\nu_e\to\nu_\tau$
transitions driven by $\Delta{m}^2_{21}$.
The occurrence of significant
$\nu_\mu\leftrightarrows\nu_e$
and
$\nu_\mu\to\nu_\tau$
transitions due to $\Delta{m}^2_{21}$
is excluded by the absence of an additional deficit of
low-energy upward-going $\mu$-like events
with respect to the main deficit due to
$\nu_\mu\to\nu_\tau$ oscillations driven by
$\Delta{m}^2_{31}$
(see Fig.3 of Ref.\cite{SK-atm-98}).
The occurrence of $\nu_e\to\nu_\tau$
transitions is excluded by the absence of
a deficit of low-energy upward-going $e$-like events
(see Fig.3 of Ref.\cite{SK-atm-98}).
Let us notice that the introduction
in the neutrino mixing scheme
of additional sterile neutrinos
(see \cite{BGW98} and references therein)
required to explain also the indications in favor of
$\bar\nu_\mu\to\bar\nu_e$
and
$\nu_\mu\to\nu_e$
transitions
found in the LSND experiment \cite{LSND}
makes things even worse,
because active neutrinos can have additional transitions into sterile states,
leading to additional unobserved deficits of both
$e$-like and $\mu$-like events.

In this paper we will present a quantitative proof of the fact that
the high-$\Delta{m}^2$ part of
the large mixing angle MSW solution of the solar neutrino problem
is disfavored by the Super-Kamiokande atmospheric neutrino data
assuming the scheme of neutrino mixing indicated
by the results of solar and atmospheric neutrino experiments
together with the result of the long-baseline experiment reactor
neutrino oscillation experiment CHOOZ \cite{CHOOZ98}.
In this scheme the element $U_{e3}$ of the neutrino mixing matrix
is negligibly small \cite{Vissani97,BG98-dec}
and the transition probabilities of atmospheric neutrinos
can be calculated analytically.
We will show that in the scheme under consideration
the high-$\Delta{m}^2$ part of
the large mixing angle MSW solution of the solar neutrino problem
is disfavored because it implies a depletion of
atmospheric upward-going low-energy $e$-like events,
which is in contradiction with the observations of the Super-Kamiokande experiment.

Our starting point is the fact that under the hypothesis of three-flavour neutrino
oscillations,
the number of $e$-like events, $N_e^{\mathrm{DATA}}$,
and the number of $\mu$-like events, $N_\mu^{\mathrm{DATA}}$,
measured in the Super-Kamiokande experiment
are given by
\begin{eqnarray}
&&
N_e^{\mathrm{DATA}}
=
N_e^{\mathrm{MC}}
+
\left( N_\mu^{\mathrm{MC}} - N_e^{\mathrm{MC}} \right) P_{e\mu}
- N_e^{\mathrm{MC}} \, P_{e\tau}
\,,
\label{51}
\\
&&
N_\mu^{\mathrm{DATA}}
=
N_\mu^{\mathrm{MC}}
-
\left( N_\mu^{\mathrm{MC}} - N_e^{\mathrm{MC}} \right) P_{e\mu}
- N_\mu^{\mathrm{MC}} \, P_{\mu\tau}
\,,
\label{52}
\end{eqnarray}
where $N_e^{\mathrm{MC}}$ and $N_\mu^{\mathrm{MC}}$
are the number of events calculated with the Monte Carlo method,
without neutrino oscillations,
and
$P_{\alpha\beta}$
is the probability of
$\nu_\alpha\to\nu_\beta$
transitions properly averaged
over the neutrino energy spectrum, the neutrino propagation distance,
and the relative amounts of neutrinos and antineutrinos.
Here we assume that,
at least after these averagings,
the probabilities of
$\nu_\alpha\to\nu_\beta$
and
$\nu_\beta\to\nu_\alpha$
transitions are equal.

\section{The neutrino mixing scheme}
\label{The neutrino mixing scheme}

In the scheme with three-neutrino mixing where $\Delta{m}^2_{21}$ is
responsible for solar neutrino oscillations
and
$\Delta{m}^2_{31}\gg\Delta{m}^2_{21}$ is 
responsible for atmospheric neutrino oscillations,
the negative result
of the CHOOZ experiment
in the search for long-baseline
$\bar\nu_e\to\bar\nu_e$
oscillations implies that
\begin{equation}
\sin^22\vartheta_{\mathrm{CHOOZ}} \leq 0.18
\quad \mbox{for} \quad
\Delta{m}^2_{\mathrm{31}} \gtrsim 2 \times 10^{-3} \, \mathrm{eV}^2
\,.
\label{451}
\end{equation}
Hence,
this limit applies to the range (\ref{406}) of $\Delta{m}^2_{\mathrm{31}}$
under consideration.

In the three-neutrino scheme with $\Delta{m}^2_{31}\gg\Delta{m}^2_{21}$
we have
\begin{equation}
\sin^22\vartheta_{\mathrm{CHOOZ}} = 4 |U_{e3}|^2 ( 1 - |U_{e3}|^2 )
\label{452}
\end{equation}
(see \cite{BGK96-atm,GKM98-atm}),
where $U$ is the $3\times3$ neutrino mixing matrix.
The upper bound (\ref{451}) on $\sin^22\vartheta_{\mathrm{CHOOZ}}$
implies that
\begin{equation}
|U_{e3}|^2 \leq 5 \times 10^{-2}
\quad \mbox{or} \quad
|U_{e3}|^2 \geq 0.95
\,.
\label{453}
\end{equation}
A large value of $|U_{e3}|^2$
fails to to explain the solar neutrino problem
with neutrino oscillations
because
in the scheme under consideration
the averaged survival probability of solar electron neutrinos
is given by \cite{Shi-Schramm92}
\begin{equation}
P_{\nu_e\to\nu_e}^{\mathrm{sun}}(E)
=
\left(
1
-
|U_{e3}|^2
\right)^2
P_{\nu_e\to\nu_e}^{(1,2)}(E)
+
|U_{e3}|^4
\,,
\label{35}
\end{equation}
where
$E$ is the neutrino energy
and
$ P_{\nu_e\to\nu_e}^{(1,2)}(E) $
is the survival probability of solar $\nu_{e}$'s
due to the mixing of $\nu_e$ with $\nu_1$ and $\nu_2$.
The expression (\ref{35}) implies that
$ P_{\nu_e\to\nu_e}^{\mathrm{sun}}(E) \geq |U_{e3}|^4 $
and
for
$ |U_{e3}|^2 \geq 0.95 $
we have
$ P_{\nu_e\to\nu_e}^{\mathrm{sun}}(E) \geq 0.90 $.
With such a high and practically constant value of
$P_{\nu_e\to\nu_e}^{\mathrm{sun}}(E)$
it is not possible to explain the suppression of the solar $\nu_e$ flux
measured by all experiments
(Homestake \cite{Homestake98},
Kamiokande \cite{Kam-sun-96},
GALLEX \cite{GALLEX96},
SAGE \cite{SAGE96}
and
Super-Kamiokande \cite{SK-sun-98-PRL,SK-sun-nu98})
with respect to that predicted by the Standard Solar Model
(see \cite{Bahcall-nu98} and references therein).
Therefore,
the results of the CHOOZ experiment together with the
results of solar and atmospheric neutrino experiments, imply that
$|U_{e3}|^2$ is small:
\begin{equation}
|U_{e3}|^2 \leq 5 \times 10^{-2}
\,.
\label{041}
\end{equation}
Such a small value of
$|U_{e3}|^2$
implies that
the oscillations of solar and atmospheric neutrinos are decoupled
\cite{BG98-dec}
and the two-generation analyses of the solar neutrino data
yield correct information on the values of
$\Delta{m}^2_{21}$
and
$ \sin^22\vartheta_{\mathrm{sun}} \simeq |U_{e2}| $.

Furthermore,
it has been shown in Ref.~\cite{BG98-dec}
that if not only $|U_{e3}|^2$
is small,
but also
$|U_{e3}|\ll1$,
the two-generation analyses of the atmospheric neutrino data
with $\nu_\mu\to\nu_\tau$ oscillations
yield correct information on the values of
$\Delta{m}^2_{31}$
and
$ \sin^22\vartheta_{\mathrm{atm}} = 4 |U_{\mu3}|^2 ( 1 - |U_{\mu3}|^2 ) $.
Hence,
in the following we will assume that
$|U_{e3}|\ll1$
\cite{Vissani97,BG98-dec}.
In this case
the neutrino mixing matrix can be written as
\begin{equation}
U
\simeq
\left(
\begin{array}{ccc}
\cos\vartheta_{\mathrm{sun}}
&
\sin\vartheta_{\mathrm{sun}}
&
0
\\
-
\sin\vartheta_{\mathrm{sun}}
\cos\vartheta_{\mathrm{atm}}
&
\cos\vartheta_{\mathrm{sun}}
\cos\vartheta_{\mathrm{atm}}
&
\sin\vartheta_{\mathrm{atm}}
\\
\sin\vartheta_{\mathrm{sun}}
\sin\vartheta_{\mathrm{atm}}
&
-
\cos\vartheta_{\mathrm{sun}}
\sin\vartheta_{\mathrm{atm}}
&
\cos\vartheta_{\mathrm{atm}}
\end{array}
\right)
\,.
\label{04}
\end{equation}
A particular case of this mixing matrix is obtained for
$ \vartheta_{\mathrm{sun}} = \vartheta_{\mathrm{atm}} = \pi/4 $
and corresponds to the
bi-maximal mixing
that has been assumed recently by several authors \cite{bi-maximal}.
Notice, however, that bi-maximal mixing
is not compatible with the large mixing angle MSW solution
of the solar neutrino problem
\cite{Giunti98-bimax}.

The assumption
$|U_{e3}|\ll1$
implies that CP and T violation in the lepton sector is negligibly small.
Indeed,
the CP-violating phase that is present
in the general expression of the mixing matrix
(see \cite{PDG98})
can be eliminated if one of the elements of the mixing matrix is zero
(this can also be seen
by noticing that in this case the Jarlskog rephasing-invariant parameter
\cite{Jarlskog-DGW-Dunietz}
is equal to zero).
Hence,
the probability of $\nu_\alpha\to\nu_\beta$
transitions in vacuum is the same as that of
$\bar\nu_\alpha\to\bar\nu_\beta$
transitions,
but
the probability of $\nu_\alpha\to\nu_\beta$
transitions in matter is different from that of
$\bar\nu_\alpha\to\bar\nu_\beta$
transitions
because neutrino and antineutrinos have different interactions
with the medium,
which induce different effective potentials.
On the other hand,
the probabilities of
$
\stackrel{\makebox[0pt][l]
{$\hskip-3pt\scriptscriptstyle(-)$}}{\nu_{\alpha}}
\to
\stackrel{\makebox[0pt][l]
{$\hskip-3pt\scriptscriptstyle(-)$}}{\nu_{\beta}}
$
and
$
\stackrel{\makebox[0pt][l]
{$\hskip-3pt\scriptscriptstyle(-)$}}{\nu_{\beta}}
\to
\stackrel{\makebox[0pt][l]
{$\hskip-3pt\scriptscriptstyle(-)$}}{\nu_{\alpha}}
$
transitions in vacuum and in matter are the same
because T is conserved in the lepton sector
and the matter distribution along a neutrino path crossing the Earth
is (approximately) symmetrical.
For simplicity,
in the following we will neglect the matter effect
for the oscillations of atmospheric neutrinos
and we will not make a distinction between neutrinos and antineutrinos.
The contribution of the matter effect
will be discussed elsewhere \cite{GKLN-98-progress}.

The probability of
$\nu_\alpha\to\nu_\beta$
transitions in vacuum is given by
\begin{equation}
P_{\nu_\alpha\to\nu_\beta}
=
\left|
U_{{\alpha}1} \, U_{{\beta}1}
+
U_{{\alpha}2} \, U_{{\beta}2}
\,
\exp\left( - i \, \frac{ \Delta{m}^2_{21} \, L }{ 2 E } \right)
+
U_{{\alpha}3} \, U_{{\beta}3}
\,
\exp\left( - i \, \frac{ \Delta{m}^2_{31} \, L }{ 2 E } \right)
\right|^2
\,.
\label{08}
\end{equation}
Let us consider upward-going low-energy neutrinos with
\begin{equation}
0.1 \, \mathrm{GeV}
\lesssim
E
\lesssim
1 \, \mathrm{GeV}
\,,
\qquad \qquad
10^3 \lesssim L \lesssim 10^4 \, \mathrm{km}
\,.
\label{301}
\end{equation}
Taking into account the value (\ref{406}) for $\Delta{m}^2_{31}$,
we have
\begin{equation}
\frac{ \Delta{m}^2_{31} \, L }{ E } \gtrsim 10
\,.
\label{07}
\end{equation}
These large values of
$ \Delta{m}^2_{31} L / 2 E $
imply that the oscillations generated by
the mass-squared difference $\Delta{m}^2_{31}$
are very rapid so that $\sin^2 \Delta{m}^2_{31} L / 2 E$ can be set to 1/2. 
Therefore, the measured probability is given by the average
of the expression (\ref{08}) over the fast oscillations
due to $\Delta{m}^2_{31}$:
\begin{equation}
P_{\nu_\alpha\to\nu_\beta}
=
\left|
U_{{\alpha}1} \, U_{{\beta}1}
+
U_{{\alpha}2} \, U_{{\beta}2}
\,
\exp\left( - i \, \frac{ \Delta{m}^2_{21} \, L }{ 2 E } \right)
\right|^2
+
U_{{\alpha}3}^2 \, U_{{\beta}3}^2
\,,
\label{09}
\end{equation}
which can be written as
\begin{equation}
P_{\nu_\alpha\to\nu_\beta}
=
\left( 1 - U_{{\alpha}3}^2 \right)
\left( 1 - U_{{\beta}3}^2 \right)
P^{(1,2)}_{\nu_\alpha\to\nu_\beta}
+
U_{{\alpha}3}^2 \, U_{{\beta}3}^2
\,,
\label{10}
\end{equation}
where
\begin{eqnarray}
P^{(1,2)}_{\nu_\alpha\to\nu_\beta}
\null & \null = \null & \null
\left( \frac{ U_{{\alpha}1}^2 }{ 1 - U_{{\alpha}3}^2 } \right)
\left(\frac{ U_{{\beta}1}^2 }{ 1 - U_{{\beta}3}^2 } \right)
+
\left(\frac{ U_{{\alpha}2}^2 }{ 1 - U_{{\alpha}3}^2 } \right)
\left(\frac{ U_{{\beta}2}^2 }{ 1 - U_{{\beta}3}^2 } \right)
\nonumber
\\
\null & \null \null & \null
+
2
\left(\frac{ U_{{\alpha}1} \, U_{{\alpha}2} }{ 1 - U_{{\alpha}3}^2 } \right)
\left(\frac{ U_{{\beta}1} \, U_{{\beta}2} }{ 1 - U_{{\beta}3}^2 } \right)
\cos\frac{ \Delta{m}^2_{21} \, L }{ 2 E }
\label{11}
\end{eqnarray}
is the probability of
$\nu_\alpha\to\nu_\beta$
transitions due to the mixing of
$\nu_\alpha$ and $\nu_\beta$
with
$\nu_1$ and $\nu_2$.
From Eq.(\ref{04}) we have
\begin{equation}
\frac{ U_{e1}^2 }{ 1 - U_{e3}^2 }
=
\cos^2\vartheta_{\mathrm{sun}}
\,,
\quad
\frac{ U_{e2}^2 }{ 1 - U_{e3}^2 }
=
\sin^2\vartheta_{\mathrm{sun}}
\,,
\quad
\frac{ U_{e1} \, U_{e2} }{ 1 - U_{e3}^2 }
=
\frac{1}{2} \, \sin2\vartheta_{\mathrm{sun}}
\label{12}
\end{equation}
and
\begin{equation}
\frac{ U_{\alpha1}^2 }{ 1 - U_{\alpha3}^2 }
=
\sin^2\vartheta_{\mathrm{sun}}
\,,
\quad
\frac{ U_{\alpha2}^2 }{ 1 - U_{\alpha3}^2 }
=
\cos^2\vartheta_{\mathrm{sun}}
\,,
\quad
\frac{ U_{\alpha1} \, U_{\alpha2} }{ 1 - U_{\alpha3}^2 }
=
- \frac{1}{2} \, \sin2\vartheta_{\mathrm{sun}}
\label{13}
\end{equation}
for $\alpha=\mu,\tau$.
Hence,
the probabilities
$P^{(1,2)}_{\nu_\alpha\to\nu_\beta}$
have a two-generation form and depend only on the two parameters
relevant for solar neutrino oscillations,
$\Delta{m}^2_{21}$
and
$\vartheta_{\mathrm{sun}}$:
\begin{equation}
P^{(1,2)}_{\nu_\alpha\to\nu_\alpha}
=
P^{(1,2)}_{\nu_\mu\leftrightarrows\nu_\tau}
=
1 - P_{21}
\,,
\qquad
P^{(1,2)}_{\nu_e\leftrightarrows\nu_\beta}
=
P_{21}
\qquad
( \alpha=e,\mu,\tau ; \ \beta=\mu,\tau )
\,,
\label{211}
\end{equation}
with
\begin{equation}
P_{21}
=
\frac{1}{2} \, \sin^22\vartheta_{\mathrm{sun}}
\left( 1 - \cos\frac{ \Delta{m}^2_{21} \, L }{ 2 E } \right)
\,.
\label{212}
\end{equation}
The complete expressions for the transition probabilities
are\footnote{We would like to thank A.Yu. Smirnov for noticing a mistake
in the expression of
$P_{\nu_e\leftrightarrows\nu_\tau}$
presented in the first version of this paper.}:
\begin{eqnarray}
&&
P_{\nu_e\to\nu_e}
=
1 - P_{21}
\,,
\label{17}
\\
&&
P_{\nu_\mu\to\nu_\mu}
=
1
-
\frac{1}{2} \, \sin^2 2\vartheta_{\mathrm{atm}}
-
\cos^4\vartheta_{\mathrm{atm}}
\,
P_{21}
\,,
\label{18}
\\
&&
P_{\nu_\tau\to\nu_\tau}
=
1
-
\frac{1}{2} \, \sin^2 2\vartheta_{\mathrm{atm}}
-
\sin^4\vartheta_{\mathrm{atm}}
\,
P_{21}
\,,
\label{19}
\\
&&
P_{\nu_e\leftrightarrows\nu_\mu}
=
\cos^2\vartheta_{\mathrm{atm}}
\,
P_{21}
\,,
\label{20a}
\\
&&
P_{\nu_e\leftrightarrows\nu_\tau}
=
\sin^2\vartheta_{\mathrm{atm}}
\,
P_{21}
\,,
\label{20b}
\\
&&
P_{\nu_\mu\leftrightarrows\nu_\tau}
=
\frac{1}{4} \, \sin^22\vartheta_{\mathrm{atm}}
\left( 2 - P_{21} \right)
\,.
\label{21}
\end{eqnarray}

It is important to notice that the probability of
$\nu_e\leftrightarrows\nu_\mu$
and
$\nu_e\leftrightarrows\nu_\tau$
transitions
are approximately equal if
$\sin^22\vartheta_{\mathrm{atm}}\simeq1$,
as indicated by the
$\nu_\mu\to\nu_\tau$
fit of all the atmospheric contained events and upward-going muons
measured in the Super-Kamiokande experiment
\cite{Suzuki-WIN99}.
In this case,
if $P_{21}\neq0$
the disappearance of electron neutrinos due to
$\nu_e\to\nu_\tau$
transitions
cannot be compensated by
$\nu_e\leftrightarrows\nu_\mu$
oscillations.
This can be seen by examining the ratio
\begin{equation}
\frac{ N_e^{\mathrm{DATA}} }{ N_e^{\mathrm{MC}} }
=
1
-
\left( 1 - R_{\mu/e}^{\mathrm{MC}} \right) P_{e\mu}
-
P_{e\tau}
\,,
\label{181}
\end{equation}
where
$ R_{\mu/e}^{\mathrm{MC}} \equiv N_\mu^{\mathrm{MC}} / N_e^{\mathrm{MC}} $.
For
$\sin^22\vartheta_{\mathrm{atm}}\simeq1$
we have
$
P_{e\tau}
\simeq
P_{e\mu}
\simeq
\frac{1}{2} \, \langle P_{21} \rangle
$,
where
$ \langle P_{21} \rangle $
indicates the average of
$P_{21}$
over the energy spectrum and propagation distance of atmospheric neutrinos.
It follows that
\begin{equation}
\frac{ N_e^{\mathrm{DATA}} }{ N_e^{\mathrm{MC}} }
\simeq
1
-
\frac{1}{2}
\left( 2 - R_{\mu/e}^{\mathrm{MC}} \right)
\langle P_{21} \rangle
\,.
\label{182}
\end{equation}
Since $ R_{\mu/e}^{\mathrm{MC}} < 2 $
for low-energy events in the Super-Kamiokande experiment,
one can see that $N_e^{\mathrm{DATA}}$
can only \emph{decrease}
with respect to
$N_e^{\mathrm{MC}}$
if
$\sin^22\vartheta_{\mathrm{atm}}\simeq1$.

The expression of
$ N_e^{\mathrm{DATA}} / N_e^{\mathrm{MC}} $
valid for any value of
$\sin^22\vartheta_{\mathrm{atm}}$
is
\begin{equation}
\frac{ N_e^{\mathrm{DATA}} }{ N_e^{\mathrm{MC}} }
=
1 - \left( 1 - R_{\mu/e}^{\mathrm{MC}} \cos^2\vartheta_{\mathrm{atm}} \right)
\langle P_{21} \rangle
\,.
\label{183}
\end{equation}
Therefore,
one can see that a non-zero value of
$ \langle P_{21} \rangle $
implies that
$N_e^{\mathrm{DATA}}$
should be smaller than
$N_e^{\mathrm{MC}}$
if
$
\sin^2\vartheta_{\mathrm{atm}}
>
(R_{\mu/e}^{\mathrm{MC}})^{-1}
$.

\section{Constraint on the large mixing angle MSW solution of the solar neutrino problem}
\label{Constraint on the large mixing angle MSW solution of the solar neutrino problem}

An experimental lower limit
\begin{equation}
\frac{ N_e^{\mathrm{DATA}} }{ N_e^{\mathrm{MC}} }
\geq
R_e^{\mathrm{min}}
\label{190}
\end{equation}
relative to low-energy upward-going $e$-like atmospheric neutrino events 
allows to constraint the value of
$\langle P_{21} \rangle$.
Indeed,
using Eq.(\ref{183}) we have
\begin{equation}
\langle P_{21} \rangle
\leq
\frac
{ 1 - R_e^{\mathrm{min}} }
{ 1 - R_{\mu/e}^{\mathrm{MC}} \cos^2\vartheta_{\mathrm{atm}} }
\,.
\label{187}
\end{equation}
This implies that
only the region in the
$\sin^22\vartheta_{\mathrm{sun}}$--$\Delta{m}^2_{21}$
such that
\begin{equation}
\sin^22\vartheta_{\mathrm{sun}}
\leq
\frac
{ 2 \left( 1 - R_e^{\mathrm{min}} \right) }
{ \left( 1 - R_{\mu/e}^{\mathrm{MC}} \cos^2\vartheta_{\mathrm{atm}} \right)
\left( 1 - \left\langle \cos\frac{ \Delta{m}^2_{21} \, L }{ 2 E } \right\rangle \right) }
\label{188}
\end{equation}
is allowed.
The brackets around the cosine indicate an appropriate average over energy and distance.

The Super-Kamiokande data relative to upward-going
($ \cos\theta < - 0.2$, where $\theta$ is the zenith angle)
$e$-like and $\mu$-like events
with momentum $ p < 0.4 \mathrm{GeV} $ are \cite{SK-atm-98}:
\begin{eqnarray}
&&
N_e^{\mathrm{DATA}} = 272 \pm 23
\,,
\label{nelda}
\\
&&
N_e^{\mathrm{MC}} = 247 \pm 10
\,,
\label{nelmc}
\\
&&
N_\mu^{\mathrm{DATA}} = 183 \pm 19
\,,
\label{nmuda}
\\
&&
N_\mu^{\mathrm{MC}} = 313 \pm 10
\,,
\label{nmumc}
\end{eqnarray}
where
the MC expected fluxes have been rescaled by a factor 1.158
with respect to the fluxes calculated in Ref.~\cite{Honda95}
(which predict
$ N_e^{\mathrm{MC}} = 214 \pm 8 $
and
$ N_\mu^{\mathrm{MC}} = 270 \pm 9 $),
as required by
the fit of all the Super-Kamiokande data with $\nu_\mu\to\nu_\tau$
oscillations
\cite{SK-atm-98}.

The ratio $R_{\mu/e}^{\mathrm{MC}}$ of expected $\mu$-like and $e$-like events
is given by
\begin{equation}
R_{\mu/e}^{\mathrm{MC}}
\equiv
\frac{ N_\mu^{\mathrm{MC}} }{ N_e^{\mathrm{MC}} } = 1.27 \pm 0.06
\,.
\label{71}
\end{equation}
The error of the theoretical calculation is about 5\%
and will be neglected in the following approximate calculations.
Since $(R_{\mu/e}^{\mathrm{MC}})^{-1}=0.79$,
as discussed at the end of the previous Section,
a value of $\langle P_{21} \rangle$
bigger than zero
implies that
$ N_e^{\mathrm{DATA}} $
should be smaller than
$ N_e^{\mathrm{MC}} $
if
$
\sin^2\vartheta_{\mathrm{atm}}
>
0.21
$,
which is practically guaranteed to be certain
by the result of the fit of all Super-Kamiokande atmospheric neutrino data
with $\nu_\mu\to\nu_\tau$ transitions
\cite{SK-atm-98,Suzuki-WIN99}.

From the data in Eqs.(\ref{nelda}) and (\ref{nelmc}),
the value of the ratio (\ref{190}) is
\begin{equation}
\frac{ N_e^{\mathrm{DATA}} }{ N_e^{\mathrm{MC}} }
=
1.10 \pm 0.10
\geq
0.97
\qquad
\mbox{at 90\% CL}
\,.
\label{191}
\end{equation}
Therefore,
we consider
\begin{equation}
R_e^{\mathrm{min}}
=
0.97
\,.
\label{192}
\end{equation}

If we consider now the fit of all Super-Kamiokande data
with neutrino oscillations,
the value of
$ \sin^2 2\vartheta_{\mathrm{atm}} $
is constrained to be bigger than about 0.90
at 90\% CL
\cite{Suzuki-WIN99}.
Hence, we have
\begin{equation}
0.34 \lesssim \cos^2\vartheta_{\mathrm{atm}} \lesssim 0.66
\,.
\label{193}
\end{equation}

Inserting the values
(\ref{71}), (\ref{192}) and (\ref{193})
in the inequality (\ref{188}) we obtain
\begin{equation}
\sin^22\vartheta_{\mathrm{sun}}
\leq
\frac{ 0.37 }{ 1 - \left\langle \cos\frac{ \Delta{m}^2_{21} \, L }{ 2 E } \right\rangle }
\,.
\label{194}
\end{equation}
The corresponding exclusion curve
in the $\sin^22\vartheta_{\mathrm{sun}}$--$\Delta{m}^2_{21}$
plane
obtained with an average of
$ \cos\frac{ \Delta{m}^2_{21} \, L }{ 2 E } $
over a constant energy spectrum in the interval
$ 100 \, \mathrm{MeV} \leq E \leq 1 \, \mathrm{GeV} $
and a constant angular distribution
$ -1 \leq \cos\theta \leq -0.2 $,
where $\theta$ is the zenith angle,
is shown in Fig.~\ref{lma}
(solid curve, the region on the right is forbidden).
The dotted line in Fig.~\ref{lma}
represents the exclusion curve obtained for
$ \sin^2 2\vartheta_{\mathrm{atm}} = 1 $
(the corresponding value of the numerator in Eq.(\ref{194})
is 0.16).
The light and dark shadowed areas in Fig.~\ref{lma} represent the 99\% CL
allowed region of the
large mixing angle MSW solution of the solar neutrino problem
obtained in Ref.~\protect\cite{Bahcall-Krastev-Smirnov98}
from the fit of the total rates of all the solar neutrino experiments
(Fig.~2).
The region below the dashed line is excluded from the day-night asymmetry measured in 
the Super-Kamiokande experiment \protect\cite{SK-sun-98-PRL,SK-sun-nu98}
(Fig.~10b of Ref.~\protect\cite{Bahcall-Krastev-Smirnov98})
and the dark shadowed region is the 99\% CL
allowed region obtained
in Ref.~\protect\cite{Bahcall-Krastev-Smirnov98}
(Fig.~10b)
from the fit of the 
rates of all solar neutrino experiments plus the day-night asymmetry measured
in the Super-Kamiokande experiment.

Let us emphasize that the limit (\ref{194})
has been obtained under several approximations
whose 
validity   cannot rigorously be proved without  more precise calculations
\cite{GKLN-98-progress}.
Hence,
the exclusion curve in Fig.~\ref{lma}
must be considered as an indication of disfavoring
the high-$\Delta{m}^2_{21}$ part of the allowed region
for the large mixing angle MSW solution of the solar neutrino problem.
From Fig.~\ref{lma}
one can see that the solar plus atmospheric data of the Super-Kamiokande experiment,
taking into account the non-observation of a day-night asymmetry
in the solar data,
disfavor practically all the allowed region
of the large mixing angle MSW solution of the solar neutrino problem.

\section{Conclusions}
\label{Conclusions}

We have considered the scheme with mixing of three massive neutrinos
indicated by the results of atmospheric and solar neutrino experiments,
together with the negative result of the CHOOZ reactor long-baseline experiment.
In this scheme the element $U_{e3}$ of the neutrino mixing matrix is negligibly
small.

\begin{figure}[ht]
\begin{center}
\mbox{\epsfig{file=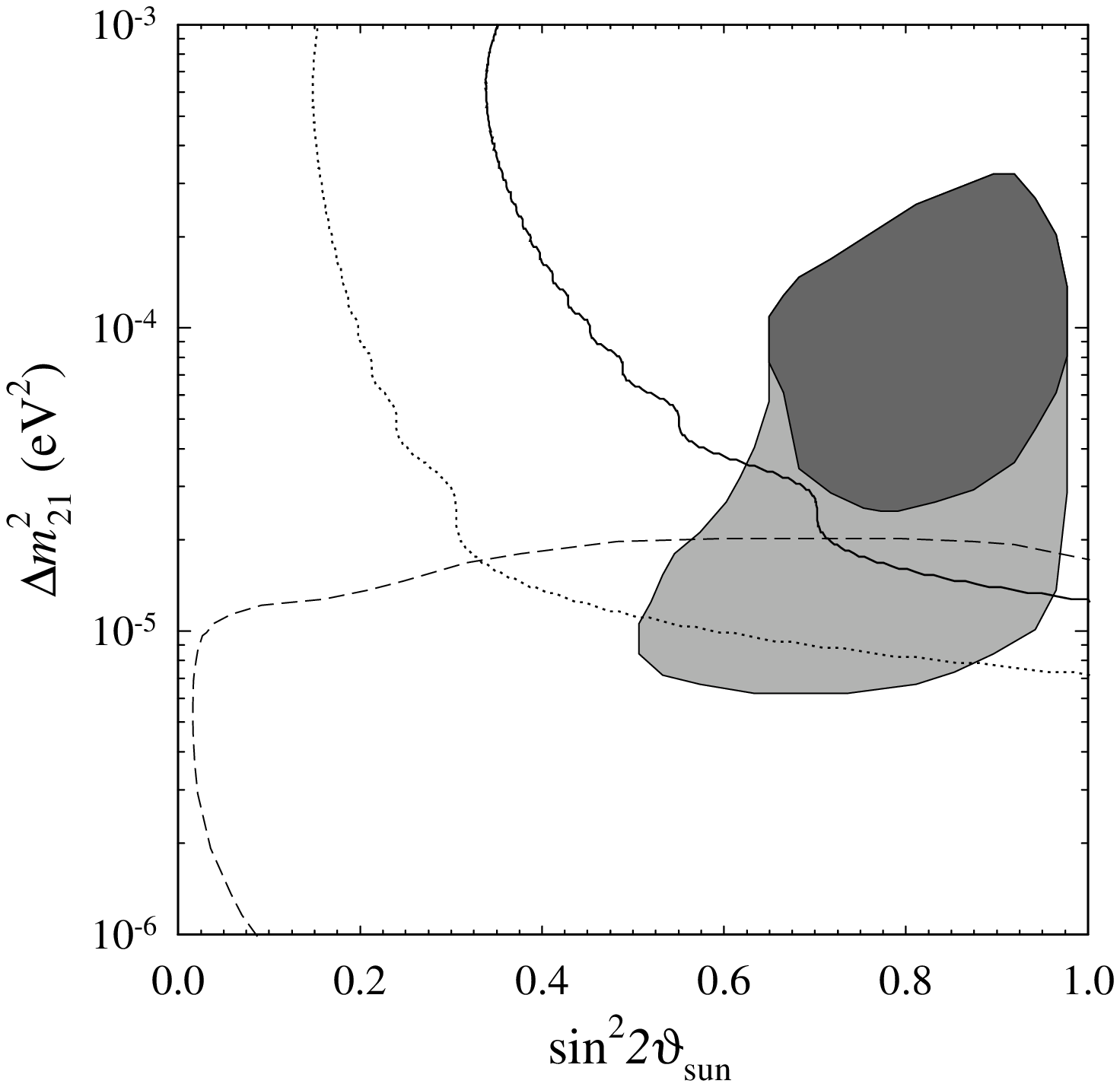,height=15cm}}
\end{center}
\caption{ \label{lma}
Exclusion curve in the $\sin^22\vartheta_{\mathrm{sun}}$--$\Delta{m}^2_{21}$
plane obtained from the Super-Kamiokande data
relative to low-energy upward-going $e$-like events
(solid line, the region on the right is forbidden).
The dotted line
represents the exclusion curve for
$ \sin^2 2\vartheta_{\mathrm{atm}} = 1 $.
The light and dark shadowed areas represent the 99\% CL
allowed region of the
large mixing angle MSW solution of the solar neutrino problem
obtained from the fit of the total rates of all solar neutrino experiments
(taken from Fig.~2 of Ref.~\protect\cite{Bahcall-Krastev-Smirnov98}).
The region below the dashed line is excluded from the day-night asymmetry measured in 
the Super-Kamiokande experiment \protect\cite{SK-sun-98-PRL,SK-sun-nu98}
(Fig.~10b of Ref.~\protect\cite{Bahcall-Krastev-Smirnov98})
and the dark shadowed region is the 99\% CL
allowed region obtained
in Ref.~\protect\cite{Bahcall-Krastev-Smirnov98}
(Fig.~10b)
from the fit of the 
rates of all solar neutrino experiments plus the day-night asymmetry measured
in the Super-Kamiokande experiment.}
\end{figure}

We have shown that if the scheme under consideration
is realized in nature,
the high-$\Delta{m}^2$ values of
the large mixing angle MSW solution of the solar neutrino problem
lead to a deficit
of low-energy upward-going atmospheric $e$-like events
with respect to the $\nu_\mu\to\nu_\tau$
fit of the experimental data.
Since this deficit has not been observed in the Super-Kamiokande experiment,
for on the contrary measured a slight excess of
low-energy upward-going atmospheric $e$-like events,
we conclude that the high-$\Delta{m}^2$ part of the large mixing angle MSW solution
of the solar neutrino problem is disfavored by the atmospheric neutrino
data of the Super-Kamiokande experiment.
We have presented in Fig.~\ref{lma}
an approximate exclusion curve obtained from the Super-Kamiokande data
which shows that the high-$\Delta{m}^2$ part of
the large mixing angle MSW solution of the solar neutrino problem
is disfavored.
Taking into account the fact that the
small-$\Delta{m}^2$ part of
the large mixing angle MSW solution of the solar neutrino problem
is excluded by the absence of a day-night asymmetry
in the Super-Kamiokande solar neutrino data,
we conclude that
the large mixing angle MSW solution of the solar neutrino problem
is disfavored by the solar and atmospheric neutrino data
of the Super-Kamiokande experiment.

As discussed in Section \ref{Introduction},
we think that the large mixing angle MSW solution
of the solar neutrino problem is disfavored by the
data of the Super-Kamiokande experiment in any scheme of neutrino mixing.
It would be very interesting to check this conclusion
performing a combined fit of the atmospheric and solar neutrino data
in the general framework of three-neutrino mixing.

It is worthwhile to emphasize that the large mixing angle MSW solution
of the solar neutrino problem
is also disfavored by the global fit of solar neutrino data,
which includes the total
rates measured in solar neutrino experiments
and the energy spectrum and zenith angle distribution of recoil electrons
measured in the Super-Kamiokande experiment
(see Fig.~15(b) of Ref.\cite{Bahcall-Krastev-Smirnov98}).

A more detailed analysis of the atmospheric neutrino data will be presented elsewhere
\cite{GKLN-98-progress}.

\acknowledgements

C.G., U.W.L. and V.A.N. are grateful to the Korea 
Institute for Advanced Study for the hospitality
extended to them while this work was done.

\begin{thebibliography}{10}

\bibitem{SK-atm-98}
Y.~Fukuda \textit{et al.},
Super-Kamiokande Coll., Phys. Rev. Lett. \textbf{81}, 1562 (1998).

\bibitem{discovery-of-neutrino-oscillations}
B. Pontecorvo, J. Exptl. Theoret. Phys. \textbf{33}, 549 (1957) [Sov. Phys.
  JETP \textbf{6}, 429 (1958)]; J. Exptl. Theoret. Phys. \textbf{34}, 247
  (1958) [Sov. Phys. JETP \textbf{7}, 172 (1958)]; Z. Maki and M. Nakagawa and
  S. Sakata, Prog. Theor. Phys. \textbf{28}, 870 (1962); B. Pontecorvo, Zh.
  Eksp. Teor. Fiz. \textbf{53}, 1717 (1967) [Sov. Phys. JETP \textbf{26}, 984
  (1968)]; V. Gribov and B. Pontecorvo, Phys. Lett. B \textbf{28}, 493 (1969).

\bibitem{solar-neutrino-problem}
See, for example: 
J.N. Bahcall, \textit{Neutrino Physics and Astrophysics}, Cambridge University
  Press, 1989; V. Berezinsky, Invited lecture at the $25^{\mathrm{th}}$
  \textit{International Cosmic Ray Conference}, Durban, 28 July -- 8 August
  1997 [astro-ph/9710126]; J.N. Bahcall, Talk presented at \textit{Neutrino
  '98} \cite{nu98} [hep-ph/9808162]; A.Yu. Smirnov, Talk presented at
  \textit{Neutrino '98} \cite{nu98} [hep-ph/9809481].

\bibitem{Kam-atm-94}
Y.~Fukuda \textit{et al.},
Kamiokande Coll., Phys. Lett. B \textbf{335}, 237 (1994).

\bibitem{IMB95}
R.~Becker-Szendy \textit{et al.},
IMB Coll., Nucl. Phys. B (Proc. Suppl.) \textbf{38}, 331 (1995).

\bibitem{Soudan97}
W.W.M. Allison \textit{et al.},
Soudan Coll., Phys. Lett. B \textbf{391}, 491 (1997).

\bibitem{MACRO-98}
M.~Ambrosio \textit{et al.},
MACRO Coll., Phys. Lett. B \textbf{434}, 451 (1998).

\bibitem{Suzuki-WIN99}
Y. Suzuki,
Talk presented at the
$17^{\mathrm{th}}$
\textit{International Workshop on Weak Interactions and Neutrinos}
(WIN99),
24--30 January 1999,
Cape Town, South Africa.

\bibitem{Homestake98}
B.T. Cleveland \textit{et al.},
Astrophys. J. \textbf{496}, 505 (1998).

\bibitem{Kam-sun-96}
K.S. Hirata \textit{et al.},
Kamiokande Coll., Phys. Rev. Lett. \textbf{77}, 1683 (1996).

\bibitem{GALLEX96}
W.~Hampel \textit{et al.},
GALLEX Coll., Phys. Lett. B \textbf{388}, 384 (1996).

\bibitem{SAGE96}
D.N. Abdurashitov \textit{et al.},
SAGE Coll., Phys. Rev. Lett. \textbf{77}, 4708 (1996).

\bibitem{SK-sun-98-PRL}
Y.~Fukuda,
Super-Kamiokande Coll., Phys. Rev. Lett. \textbf{81}, 1158 (1998).

\bibitem{SK-sun-nu98}
Y.~Suzuki,
Super-Kamiokande Coll., Talk presented at \textit{Neutrino '98} \cite{nu98},
1998.

\bibitem{MSW}
S.P. Mikheyev and A.Yu. Smirnov,
Yad. Fiz. \textbf{42}, 1441 (1985)
[Sov. J. Nucl. Phys. \textbf{42}, 913 (1985)];
Il Nuovo Cimento C \textbf{9}, 17 (1986);
L.~Wolfenstein,
Phys. Rev. D \textbf{17}, 2369 (1978);
\textit{ibid.} \textbf{20}, 2634 (1979).

\bibitem{Bahcall-Krastev-Smirnov98}
J.N. Bahcall, P.I. Krastev and A.Yu. Smirnov,
Phys. Rev. D \textbf{58}, 096016 (1998).

\bibitem{KimLee98}
C.W. Kim and U.W. Lee,
Phys. Lett. B \textbf{444}, 204 (1998).

\bibitem{BGW98}
S.M. Bilenky, C. Giunti and W. Grimus,
preprint hep-ph/9812360.

\bibitem{LSND}
C.~Athanassopoulos \textit{et al.},
LSND Coll., Phys. Rev. Lett. \textbf{77}, 3082 (1996);
Phys. Rev. Lett. \textbf{81}, 1774 (1998).

\bibitem{CHOOZ98}
M.~Apollonio \textit{et al.},
CHOOZ Coll., Phys. Lett. B \textbf{420}, 397 (1998).

\bibitem{Vissani97}
F.~Vissani,
preprint hep-ph/9708483 (1997).

\bibitem{BG98-dec}
S.M. Bilenky and C.~Giunti,
preprint hep-ph/9802201 (1998)
[to be published in Phys. Lett. B].

\bibitem{BGK96-atm}
S.M. Bilenky, C.~Giunti and C.W. Kim,
Astrop. Phys. \textbf{4}, 241 (1996).

\bibitem{GKM98-atm}
C.~Giunti, C.W. Kim and M.~Monteno,
Nucl. Phys. B \textbf{521}, 3 (1998).

\bibitem{Shi-Schramm92}
X.~Shi and D.N. Schramm,
Phys. Lett. B \textbf{283}, 305 (1992).

\bibitem{Bahcall-nu98}
J.N. Bahcall,
Talk presented at \textit{Neutrino '98} \cite{nu98}, 1998
(hep-ph/9808162).

\bibitem{bi-maximal}
V. Barger, S. Pakvasa, T.J. Weiler and K. Whisnant,
Phys. Lett. B \textbf{437}, 107 (1998);
A.J. Baltz, A.S. Goldhaber and M. Goldhaber,
Phys. Rev. Lett. \textbf{81}, 5730 (1998);
Y. Nomura and T. Yanagida,
Phys. Rev. D \textbf{59}, 017303 (1999);
G. Altarelli and F. Feruglio,
Phys. Lett. B \textbf{439}, 112 (1998);
JHEP \textbf{9811}, 021 (1998);
E. Ma,
Phys. Lett. B \textbf{442}, 238 (1998);
N. Haba,
Phys. Rev. D \textbf{59}, 035011 (1999);
H. Fritzsch and Z.Z. Xing,
Phys. Lett. B \textbf{440}, 313 (1998);
H. Georgi and S.L. Glashow,
hep-ph/9808293;
S. Davidson and S.F. King,
Phys. Lett. B \textbf{445}, 191 (1998);
R.N. Mohapatra and S. Nussinov,
Phys. Lett. B \textbf{441}, 299 (1998);
hep-ph/9809415.

\bibitem{Giunti98-bimax}
C.~Giunti,
preprint hep-ph/9810272 (1998)
[Phys. Rev. D, in press].

\bibitem{PDG98}
C.~Caso \textit{et al.},
Particle Data Group, Eur. Phys. J. C \textbf{3}, 1 (1998).

\bibitem{Jarlskog-DGW-Dunietz}
C. Jarlskog, Phys. Rev. Lett. \textbf{55}, 1039 (1985); Z. Phys. C \textbf{29},
  491 (1985); C. Jarlskog and R. Stora, Phys. Lett. B \textbf{208}, 268 (1988);
  C. Jarlskog, Proc. of \textit{CP Violation}, Ed. C. Jarlskog, Advanced Series
  in High Energy Physics Vol. 3 (World Scientific, Singapore, 1989), p. 3; I.
  Dunietz, O.W. Greenberg and D.D. Wu, Phys. Rev. Lett. \textbf{55}, 2935
  (1985); D.D. Wu, Phys. Rev. D \textbf{33}, 860 (1986); J.D. Bjorken and I.
  Dunietz, Phys. Rev. D \textbf{36}, 2109 (1987); I. Dunietz, Ann. Phys.
  \textbf{184}, 350 (1988).

\bibitem{GKLN-98-progress}
C. Giunti, C.W. Kim, U.W. Lee and V.A. Naumov, work in progress.

\bibitem{Honda95}
M.~Honda, T.~Kajita, K.~Kasahara and S.~Midorikawa,
Phys. Rev. D \textbf{52}, 4985 (1995).

\bibitem{nu98}
International Conference on Neutrino Physics and Astrophysics \textit{Neutrino
  '98}, Takayama, Japan, June 1998; WWW page:
  http://\-www-\-sk.\-icrr.\-u-\-tokyo.\-ac.\-jp/\-nu98.

\end{thebibliography}
\end{document}